\documentclass[reprint,aps,pra,showpacs,floatfix]{revtex4-1}

\usepackage{graphicx, amsmath,amssymb,psfrag,empheq}
\usepackage{dcolumn}
\usepackage{bm}
\usepackage{epsfig}
\begin{document}

\title{Exact Solutions to Non-Linear Symmetron  Theory: \\
One and Two Mirror Systems}

\author{Philippe Brax}
\email{philippe.brax@ipht.fr}
\affiliation{Institut de Physique Th\'{e}orique, Universit\'e Paris-Saclay, CEA, CNRS, F-91191 Gif/Yvette Cedex, France}

\author{Mario Pitschmann}
\email{mario.pitschmann@tuwien.ac.at}
\affiliation{Atominstitut, Technische Universit\"at Wien, Stadionallee 2, A-1020 Wien, Austria}

\date{December 16, 2017}

\begin{abstract}
We derive the exact analytical solutions to the symmetron field theory equations in the presence of a one or two mirror system. The one dimensional equations of motion are integrated exactly for both systems and their solutions can be expressed in terms of Jacobi elliptic functions. Surprisingly, in the case of two parallel mirrors the equations of motion generically provide not a unique  solution but a discrete set of solutions with increasing number of nodes and energies. The solutions obtained herein can be applied to \textit{q}BOUNCE experiments, neutron interferometry and for the calculation of the symmetron field induced "Casimir force" in the CANNEX experiment.
\end{abstract}


\maketitle


\section{Introduction}

The accelerated expansion of the universe may require the introduction of additional degrees of freedom  (see \cite{Joyce:2014kja} for a recent review). Such new degrees of freedom, in particular light scalars, are theoretically well motivated irrespectively of their role for the acceleration of the expansion of the Universe. If they exist in Nature, they must appear in some screened form in order
to prevent detection in all past experiments and observations involving scalar fifth forces. A number of screening mechanisms exist \cite{Joyce:2014kja}, the chameleon \cite{Khoury:2003rn,Khoury:2003aq,Brax:2004qh} and Damour-Polyakov mechanisms \cite{Damour:1994zq}, the K-mouflage \cite{Babichev:2009ee,Brax:2012jr,Brax:2014wla} and Vainshtein ones \cite{Vainshtein:1972sx}, allowing
such hypothetical new fields to remain unseen in local tests of gravity.

In the Damour-Polyakov  mechanism the coupling to matter weakens in regions of high density or high  Newtonian potential.
A particular representative of this effect is the symmetron mechanism \cite{Hinterbichler:2010es, Hinterbichler:2011ca} (for earlier work see \cite{Pietroni:2005pv, Olive:2007aj}),
in which the coupling of the scalar to matter is proportional to the vacuum expectation value (VEV) of the field. The
effective potential of the scalar field is such that it acquires a nonzero VEV in low-density regions, while the symmetry is restored
in high density regions. Thus, in such high density regions, the field effectively disappears as it is screened from any observation or measurement. In
regions of low density on the other hand, the field spontaneously breaks symmetry and acquires a non-vanishing VEV, couples to matter and mediates a fifth force.

When the mass of the symmetron field in the cosmological vacuum is small, e.g. of order of $10^3 H_0$ where $H_0$ is the Hubble rate now, the symmetron field could have cosmological implications \cite{Hinterbichler:2011ca} in particular on the growth of  perturbations and large-scale structure ~\cite{Clampitt:2011mx,Taddei:2013bsk}. These effects can be captured using $N$-body simulations in order to obtain non-linear properties of  the matter power spectrum and the  halo mass function, see \cite{Davis:2011pj, Winther:2011qb}. Although the symmetron is not directly coupled to photons as its matter coupling is conformal, quantum effects can give a direct interaction to electromagnetism \cite{Brax:2010uq} which could lead to  signatures of cosmological domain walls \cite{Olive:2010vh, Olive:2012ck}.
Symmetron fields for inflation have also  been investigated in \cite{Dong:2013swa}.
In \cite{Burrage:2014oza,Burrage:2015lya,Burrage:2016rkv,Brax:2016wjk}, atomic
interferometry was used  to constrain symmetrons with masses below the dark energy scale. It was found that, whilst symmetrons with masses $m$ slightly larger
than the present Hubble rate $m \sim 10^3 H_0$  have implications cosmologically \cite{Brax:2011aw}, they cannot be tested by atomic interferometry, whereas symmetrons with masses of order of the dark energy scale $m \sim 10^{-3}$ eV are within reach of the
E\"otwash types of experiments \cite{Upadhye:2012rc}. Recent bounds on symmetrons have been obtained by Jaffe \textit{et al.} \cite{Jaffe:2016fsh}.

In \cite{Cronenberg:2017aa}, gravity resonance spectroscopy \cite{Abele:2009dw, Jenke:2011zz} is used for the first time to put new bounds on symmetrons. The experimental analysis depends heavily on the field profile of symmetrons in both the one mirror or two mirror setups, where the field is either present over an infinite plane of high density or is confined between two such parallel planes.
Here, we show that an idealized 1-dimensional setup of a single mirror covering an infinite half-space or two parallel mirrors of finite separation each covering an infinite half-space admit exact analytical solutions, which can be expressed in terms of Jacobi elliptic functions.
Moreover, we find that, very surprisingly for highly non-linear differential equations, solutions are generically not unique but rather show  a discrete spectrum of possible solutions with an increasing number of nodes and energies. For a similar study applied to chameleon field theory see \cite{Ivanov:2016rfs}.

In section \ref{sec:1} we will recall some background information on symmetrons, which will provide the relevant definitions for the field theory analysis. In section \ref{sec:2} the solutions for the one mirror case will be derived, while in sections \ref{sec:3} and \ref{sec:4} the symmetric and anti-symmetric two mirror solutions will be given and discussed in section \ref{sec:5}. As an illustration, in section \ref{sec:6} a particular case study is carried out for an arbitrary choice of parameters and the complete spectrum of solutions is derived. Section \ref{sec:7} provides relevant information on the \textit{q}BOUNCE experiment, where the symmetron induced resonance frequency shift for the case of a single mirror has also been summarized  for a large range of parameters. A conclusion in section \ref{sec:8} will be followed by an Appendix providing additional technical details on the screening of a neutron.

\section{Background}\label{sec:1}

Following \cite{Joyce:2014kja}, the symmetron potential is given by
\begin{align}
   V(\phi) = -\frac{\mu^2}{2}\,\phi^2 + \frac{\lambda}{4}\,\phi^4\>,
\end{align}
with a parameter $\mu$ of dimension mass and the dimensionless self-interaction coupling $\lambda$. Together with the coupling to matter this induces an effective potential
\begin{align}
   V_\text{eff}(\phi) = V(\phi) + A(\phi)\,\rho\>,
\end{align}
where for the symmetron we have
\begin{align}\label{cf}
   A(\phi) = 1 + \frac{\phi^2}{2M^2} + \mathcal O(\phi^4/M^4)\>,
\end{align}
and hence
\begin{align}
   V_\text{eff}(\phi) = \frac{1}{2}\left(\frac{\rho}{M^2} - \mu^2\right)\phi^2 + \frac{\lambda}{4}\,\phi^4\>.
\end{align}
Here, we have neglected an additional term $\rho$, which does not affect the equations of motion.

In the "symmetric phase" where $\rho_s \geq M^2\mu^2$ we use
\begin{align}
   \rho_\text{eff} = \rho_s - M^2\mu^2 \geq 0\>,
\end{align}
while in the "broken symmetry phase" where $\rho_b < M^2\mu^2$ we use
\begin{align}
   \mu_\text{eff}^2 = \mu^2 - \frac{\rho_b}{M^2} > 0\>.
\end{align}
For the relevant experimental situations such as the \textit{q}BOUNCE experiments, neutron interferometry or Casimir experiments like CANNEX, we typically have $\rho_\text{eff} \simeq \rho_s$ and $\mu_\text{eff}^2 \simeq \mu^2$, still the exact solutions hold for more general situations.

The 1-dimensional Hamiltonian is given by
\begin{align}\label{Ham}
   \mathcal H &= \frac{1}{2}\left(\frac{d\phi}{dz}\right)^2 + V_\text{eff}(\phi) \nonumber\\
   &= \frac{1}{2}\left(\frac{d\phi}{dz}\right)^2 + \frac{1}{2}\left(\frac{\rho}{M^2} - \mu^2\right)\phi^2 + \frac{\lambda}{4}\,\phi^4\>.
\end{align}
The two vacuum values of $\phi$, given by the equation $V_{\text{eff},\phi}(\phi)\big|_{\phi=\pm\phi_V}=0$, read $\pm\phi_V$ where
\begin{align}
   \phi_V := \sqrt{\frac{\mu_\text{eff}^2}{\lambda}}\>.
\end{align}
 For static solutions we have
\begin{align}
   \frac{d^2\phi}{dz^2} = V_{\text{eff},\phi}(\phi)\>.
\end{align}
Multiplication by $\displaystyle\frac{d\phi}{dz}$ and integration with respect to $z$ provides the important relation
\begin{align}\label{1LD1}
   \frac{1}{2}\left(\frac{d\phi}{dz}\right)^2 - \frac{1}{2}\left(\frac{d\phi}{dz}\right)^2\bigg|_{z=z_0} = V_\text{eff}(\phi) - V_\text{eff}(\phi)\big|_{z=z_0}\>,
\end{align}
which determines the solutions provided both the value of the field and its first derivative are given at a point $z_0$.

\section{1 Mirror}\label{sec:2}

In this section, we treat the case of a single mirror filling the infinite half-space $z \leq 0$.

\subsection{"Broken Symmetry Phase"}

First, we consider the case of low density $\rho_b < M^2\mu^2$  corresponding to the medium above the mirror and
search for a solution that asymptotically for $z \to \infty$ goes as $\phi(z) \to \pm\phi_V$
implying $\displaystyle\frac{d\phi}{dz} \to 0$. Without loss of generality we consider $\phi(z) \to +\phi_V$ and find for $z_0=0$ and $z\to\infty$
\begin{align}\label{1LD2}
   - \frac{1}{2}\left(\frac{d\phi}{dz}\right)^2\bigg|_{z=0} = V_\text{eff}(\phi_V) - V_\text{eff}(\phi)\big|_{z=0}\>.
\end{align}
Subtracting Eq.~(\ref{1LD2}) from (\ref{1LD1}) gives
\begin{align}
   \frac{1}{2}\left(\frac{d\phi}{dz}\right)^2 = V_\text{eff}(\phi) - V_\text{eff}(\phi_V)\>,
\end{align}
leading to
\begin{align}
   \int_{\phi_0}^{\phi(z)}\frac{d\phi}{\sqrt{-\mu_\text{eff}^2\,\big(\phi^2 - \phi_V^2\big) + \lambda/2\,\big(\phi^4 - \phi_V^4\big)}} = z\>.
\end{align}
With $\displaystyle y:= \frac{\phi(z)}{\phi_V}\>$ and $\displaystyle k:= \frac{\phi_0}{\phi_V}\>$, the latter being the ratio between the value of $\phi$ taken for $z=0$ and the vacuum value $\phi_V$, we find
\begin{align}
   z &= \frac{1}{\mu_\text{eff}}\int_k^y\frac{dy'}{\sqrt{\displaystyle1 - y'^2 + 1/2\,\big(y'^4 - 1\big)}}  \nonumber\\
   &= \frac{\sqrt2}{\mu_\text{eff}}\,\big(\tanh^{-1}y - \tanh^{-1}k\big)  \nonumber\\
   &= \frac{\sqrt2}{\mu_\text{eff}}\,\tanh^{-1}\Big(\frac{y - k}{1 - ky}\Big)\>.
\end{align}
Inverting the relation straightforwardly leads to
\begin{align}
   \phi(z) = \phi_V\frac{k}{|k|}\frac{\displaystyle |k| + \tanh\!\Big(\frac{\mu_\text{eff} z}{\sqrt2}\Big)}{\displaystyle 1 + |k|\,\tanh\!\Big(\frac{\mu_\text{eff} z}{\sqrt2}\Big)}\>,
\end{align}
or equivalently
\begin{align}
   \phi(z) = \phi_V\frac{k}{|k|}\tanh\!\Big(\frac{\mu_\text{eff} z}{\sqrt2} + \tanh^{-1}|k|\Big)\>.
\end{align}

\subsection{"Symmetric Phase"}

Here, we consider the case of high density $\rho_s \geq M^2\mu^2$, as inside the mirror.
Clearly, for $z\to-\infty$ we have $\phi(z)\to0$ and hence $\phi'\to0$. Therefore, we find for $z_0=0$ and $z\to-\infty$
\begin{align}
   - \frac{1}{2}\left(\frac{d\phi}{dz}\right)^2\bigg|_{z=0} = - V_\text{eff}(\phi)\big|_{z=0}\>.
\end{align}
Subtracting the latter equation of Eq.~(\ref{1LD1}) gives
\begin{align}
   \frac{1}{2}\left(\frac{d\phi}{dz}\right)^2 = V_\text{eff}(\phi)\>.
\end{align}
Without loss of generality we consider the positive solution $\phi(z)\geq0$. Then, for $z \leq 0$ we obviously have
\begin{align}
   \frac{d\phi}{dz} = \sqrt{\frac{\rho_\text{eff}}{M^2}\,\phi^2 + \frac{\lambda}{2}\,\phi^4}\>,
\end{align}
or with $\phi_0 := \phi(0)$
\begin{align}
   \int_{\phi_0}^{\phi(z)}\frac{d\phi}{\displaystyle\phi\,\sqrt{1 + \frac{\lambda M^2}{2\rho_\text{eff}}\,\phi^2}} = \frac{\sqrt{\rho_\text{eff}}}{M}\,z\>.
\end{align}
Since for positive $x$
\begin{align}
   \frac{d}{dx}\bigg\{-\ln\frac{1 + \sqrt{1 + a\,x^2}}{x}\bigg\} = \frac{1}{x\,\sqrt{1 + ax^2}}\>,
\end{align}
we find
\begin{align}
   \phi(z) = \frac{\phi_0}{\displaystyle\cosh\Big(\frac{\sqrt{\rho_\text{eff}}}{M}\,z\Big) - \sqrt{1 + \frac{\lambda M^2}{2\rho_\text{eff}}\,\phi_0^2}\,\sinh\Big(\frac{\sqrt{\rho_\text{eff}}}{M}\,z\Big)}\>,
\end{align}
or equivalently
\begin{align}
   &\phi(z) = \nonumber\\
   &-\sqrt{\frac{2\rho_\text{eff}}{\lambda}}\frac{1}{M}\frac{1}{\displaystyle\sinh\bigg(\frac{\sqrt{\rho_\text{eff}}}{M}\,z - \sinh^{-1}\Big(\sqrt{\frac{2\rho_\text{eff}}{\lambda}}\frac{1}{M\phi_0}\Big)\bigg)}\>.
\end{align}
This can also be expressed in terms of \textit{Jacobi elliptic functions} since
\begin{align}
   \sinh u = \textrm{sc}(u,1) = \textrm{sd}(u,1)\>.
\end{align}

\subsection{Boundary Conditions}

Using the boundary conditions
\begin{align}
   \frac{1}{2}\left(\frac{d\phi}{dz}\right)^2\bigg|_{z=0_-} = \frac{1}{2}\left(\frac{d\phi}{dz}\right)^2\bigg|_{z=0_+}\>,
\end{align}
we find
\begin{align}
  \frac{\rho_\text{eff}}{2M^2}\,\phi_0^2 + \frac{\lambda}{4}\,\phi_0^4 = - \frac{\mu_\text{eff}^2}{2}\,\big(\phi_0^2 - \phi_V^2\big) + \frac{\lambda}{4}\,\big(\phi_0^4 - \phi_V^4\big)\>,
\end{align}
or
\begin{align}
  |\phi_0| = \frac{\phi_V}{\sqrt2}\frac{1}{\displaystyle\sqrt{1 + \frac{\rho_\text{eff}}{M^2\mu_\text{eff}^2}}}\>,
\end{align}
leading to
\begin{align}
  |k| = \frac{1}{\sqrt2}\frac{1}{\displaystyle\sqrt{1 + \frac{\rho_\text{eff}}{M^2\mu_\text{eff}^2}}}\>.
\end{align}
The second boundary condition
\begin{align}
   \phi(0_-) = \phi(0_+)\>,
\end{align}
is trivially satisfied.

\subsection{Final Solution}

Finally, we obtain the solution
\begin{align}\label{FS1M}
   &\phi(z) = \Theta(+z)\,\phi_V\frac{k}{|k|}\tanh\!\Big(\frac{\mu_\text{eff} z}{\sqrt2} + \tanh^{-1}|k|\Big) - \Theta(-z) \nonumber\\
   &\times\sqrt{\frac{2\rho_\text{eff}}{\lambda}}\frac{1}{M}\frac{1}{\displaystyle\sinh\bigg(\frac{\sqrt{\rho_\text{eff}}}{M}\,z - \sinh^{-1}\Big(\sqrt{\frac{2\rho_\text{eff}}{\lambda}}\frac{1}{Mk\phi_V}\Big)\bigg)}\>,
\end{align}
where $k$ is given by
\begin{align}
  k = \pm\frac{1}{\sqrt2}\frac{1}{\displaystyle\sqrt{1 + \frac{\rho_\text{eff}}{M^2\mu_\text{eff}^2}}}\>,
\end{align}
which determines the boundary value of the scalar field.
This solution has been used in \cite{Cronenberg:2017aa} in order to evaluate the energy levels of a neutron in a \textit{q}BOUNCE experiment.

\section{2 Mirrors: Symmetric Solution}\label{sec:3}

In this section, we treat the case of two parallel infinitely thick mirrors separated at distance $2d$ in $z$-direction, with $z=0$ being the center between the two mirrors. First, we consider symmetric solutions only.

\subsection{"Broken Symmetry Phase"}

Here, we consider the case of low density $\rho_b < M^2\mu^2$, as between the mirrors and choose $z_0=0$. Due to the symmetry of the setup the derivative of the field has to vanish there yielding
\begin{align}
   \frac{1}{2}\left(\frac{d\phi}{dz}\right)^2 = V_\text{eff}(\phi) - V_\text{eff}(\phi_0)\>,
\end{align}
where $\phi_0 := \phi(0)$. Without loss of generality we take $\phi_0 > 0$. Then, for $0\leq z \leq d$ we have
\begin{align}
   \frac{d\phi}{dz} = -\sqrt{2\,\big(V_\text{eff}(\phi) - V_\text{eff}(\phi_0)\big)}\>,
\end{align}
and
\begin{align}
   \int_{\phi_0}^{\phi(z)}\frac{d\phi}{\sqrt{-\mu_\text{eff}^2\,\big(\phi^2 - \phi_0^2\big) + \lambda/2\,\big(\phi^4 - \phi_0^4\big)}} = -z\>.
\end{align}
With $\displaystyle x:= \frac{\phi(z)}{\phi_0}\>$, and taking $\mu_\text{eff} > 0$, we have
\begin{align}
   -z &= \frac{1}{\mu_\text{eff}\,\sqrt{1 - k^2/2}}\int_1^x\frac{dx'}{\displaystyle\sqrt{\big(1 - x'^2\big)\left(1 - \frac{k^2/2}{1 - k^2/2}\,x'^2\right)}}  \nonumber\\
   &= \frac{1}{\mu_\text{eff}\,\sqrt{1 - k^2/2}} \nonumber\\
   &\times \bigg\{F\bigg(\arcsin(x),\frac{k/\sqrt2}{\sqrt{1 - k^2/2}}\bigg) - F\bigg(\frac{\pi}{2},\frac{k/\sqrt2}{\sqrt{1 - k^2/2}}\bigg)\bigg\}\>,
\end{align}
where we have employed the \textit{Elliptic Integral of the first kind}
\begin{align}
   F(\phi,\ell) = \int_0^{\sin\phi}\frac{dt}{\sqrt{(1 - t^2)(1 - \ell^2t^2)}}\>.
\end{align}
Here, $k$ gives the ratio between the maximum value of the field between the two plates and the vacuum value $\phi_V$ in the absence of the two plates, i.e. the ratio $k$ captures the effect of confinement of the field between the plates.
Hence, we obtain
\begin{align}
   \phi(z) &= \phi(0)\sin\bigg\{F^{-1}\bigg[F\bigg(\frac{\pi}{2},\frac{k/\sqrt2}{\sqrt{1 - k^2/2}}\bigg) \nonumber\\
   &\quad- \mu_\text{eff}\,\sqrt{1 - k^2/2}\,z,\frac{k/\sqrt2}{\sqrt{1 - k^2/2}}\bigg]\bigg\}\>.
\end{align}
With the \textit{Jacobi elliptic functions}
\begin{align}
   \textrm{sn}(u,\ell) &= \sin\big(F^{-1}(u,\ell)\big)\>, \nonumber\\
   \textrm{cn}(u,\ell) &= \cos\big(F^{-1}(u,\ell)\big)\>, \nonumber\\
   \textrm{dn}(u,\ell) &= \sqrt{1 - \ell^2\sin^2\!\big(F^{-1}(u,\ell)\big)}\>,
\end{align}
we can write this as
\begin{align}
   \phi(z) &= \phi(0)\,\textrm{sn}\bigg\{F\bigg(\frac{\pi}{2},\frac{k/\sqrt2}{\sqrt{1 - k^2/2}}\bigg) \nonumber\\
   &\quad- \mu_\text{eff}\,\sqrt{1 - k^2/2}\,z,\frac{k/\sqrt2}{\sqrt{1 - k^2/2}}\bigg\}\>.
\end{align}
Since
\begin{align}
   \textrm{sn}\Big(u + F\Big(\frac{\pi}{2},\ell\Big),\ell\Big) = \frac{\textrm{cn}(u,\ell)}{\textrm{dn}(u,\ell)} = \textrm{cd}(u,\ell)\>,
\end{align}
and
\begin{align}
   \textrm{sn}(-u,\ell) &= -\textrm{sn}(u,\ell)\>, \nonumber\\
   \textrm{cn}(-u,\ell) &= +\textrm{cn}(u,\ell)\>, \nonumber\\
   \textrm{dn}(-u,\ell) &= +\textrm{dn}(u,\ell)\>,
\end{align}
we have, also for $-d \leq z \leq d$ and positive as well as negative $\phi(0)$
\begin{align}
   \phi(z) = \phi_Vk\,\textrm{cd}\bigg\{\mu_\text{eff}\,\sqrt{1 - k^2/2}\,z,\frac{|k|/\sqrt2}{\sqrt{1 - k^2/2}}\bigg\}\>.
\end{align}

\subsection{"Symmetric Phase"}

We can read off the solution inside the mirrors directly from the corresponding solution in the 1-mirror case
\begin{align}
   &\phi(z) = \sqrt{\frac{2\rho_\text{eff}}{\lambda}}\frac{1}{M} \nonumber\\
   &\times\frac{1}{\displaystyle\sinh\bigg(\frac{\sqrt{\rho_\text{eff}}}{M}\,(|z| - d) + \sinh^{-1}\Big(\sqrt{\frac{2\rho_\text{eff}}{\lambda}}\frac{1}{M\phi_d}\Big)\bigg)}\>,
\end{align}
where $\phi_d := \phi(d)$ and hence can be positive or negative.

\subsection{Boundary Conditions}

Using the boundary conditions at the mirror surface
\begin{align}
   \frac{1}{2}\left(\frac{d\phi}{dz}\right)^2\bigg|_{z=d_-} = \frac{1}{2}\left(\frac{d\phi}{dz}\right)^2\bigg|_{z=d_+}\>,
\end{align}
we find
\begin{align}
  -\frac{\mu_\text{eff}^2}{2}\,\big(\phi_d^2 - \phi_0^2\big) + \frac{\lambda}{4}\,\big(\phi_d^4 - \phi_0^4\big) = \frac{\rho_\text{eff}}{2M^2}\,\phi_d^2 + \frac{\lambda}{4}\,\phi_d^4\>,
\end{align}
or
\begin{align}
  |\phi_d| = \phi_V|k|\,\frac{\displaystyle\sqrt{1 - \frac{k^2}{2}}}{\displaystyle\sqrt{1 + \frac{\rho_\text{eff}}{M^2\mu_\text{eff}^2}}}\>.
\end{align}
Using this in the second boundary condition
\begin{align}
   \phi(d_-) = \phi(d_+)\>,
\end{align}
provides the relation
\begin{align}
   \sqrt{1 - \frac{k^2}{2}} &= \sqrt{1 + \frac{\rho_\text{eff}}{M^2\mu_\text{eff}^2}} \nonumber\\
   &\quad\times\bigg|\textrm{cd}\bigg\{\mu_\text{eff}\,\sqrt{1 - k^2/2}\,d,\frac{|k|/\sqrt2}{\sqrt{1 - k^2/2}}\bigg\}\bigg|\>.
\end{align}
The solution of the equation above gives possible values of $k$, i.e. it determines the value $\phi_0$ of the field between the two plates. It will give all solutions for which $\phi'(d_-) = \pm\phi'(d_+)$, hence one still has to extract the subset of solutions satisfying $\phi'(d_-) = \phi'(d_+)$.

\subsection{Final Solution}

Finally, we obtain the solution
\begin{align}
   \phi(z) &= \Theta(d - |z|)\,\phi_Vk\,\textrm{cd}\bigg\{\mu_\text{eff}\,\sqrt{1 - k^2/2}\,z,\frac{|k|/\sqrt2}{\sqrt{1 - k^2/2}}\bigg\} \nonumber\\
   &+ \Theta(|z| - d)\,\sqrt{\frac{2\rho_\text{eff}}{\lambda}}\frac{1}{M} \nonumber\\
   &\times\frac{\mathfrak{sgn_S}}{\displaystyle\sinh\bigg(\frac{\sqrt{\rho_\text{eff}}}{M}\,(|z| - d) + \sinh^{-1}\Big(\sqrt{\frac{2\rho_\text{eff}}{\lambda}}\frac{1}{M\phi_d}\Big)\bigg)}\>,
\end{align}
where
\begin{align}
  \mathfrak{sgn_S} := \frac{\textrm{cd}\bigg\{\mu_\text{eff}\,\sqrt{1 - k^2/2}\,d,\frac{|k|/\sqrt2}{\sqrt{1 - k^2/2}}\bigg\}}{\bigg|\textrm{cd}\bigg\{\mu_\text{eff}\,\sqrt{1 - k^2/2}\,d,\frac{|k|/\sqrt2}{\sqrt{1 - k^2/2}}\bigg\}\bigg|}\>,
\end{align}
$k$ is solution of
\begin{align}\label{SSS}
   \sqrt{1 - \frac{k^2}{2}} &= \sqrt{1 + \frac{\rho_\text{eff}}{M^2\mu_\text{eff}^2}} \nonumber\\
   &\quad\times\bigg|\textrm{cd}\bigg\{\mu_\text{eff}\,\sqrt{1 - k^2/2}\,d,\frac{|k|/\sqrt2}{\sqrt{1 - k^2/2}}\bigg\}\bigg|\>,
\end{align}
and
\begin{align}
  \phi_d = \pm\phi_Vk\,\frac{\displaystyle\sqrt{1 - \frac{k^2}{2}}}{\displaystyle\sqrt{1 + \frac{\rho_\text{eff}}{M^2\mu_\text{eff}^2}}}\>.
\end{align}
As mentioned already, Eq.~(\ref{SSS}) will give all solutions for which $\phi'(d_-) = \pm\phi'(d_+)$, hence one still has to extract the subset of solutions satisfying $\phi'(d_-) = \phi'(d_+)$.

Clearly, the expression on the left-hand side of Eq.~(\ref{SSS}) is $\mathcal O(1)$. The first factor on the right-hand side is for most cases of interest, where typically $\rho_\text{eff} \gg M^2\mu_\text{eff}^2$, comparably large. This implies that for a solution the second factor must be small, viz. it will be close to a zero of the \textit{Jacobi function} cd.
The Taylor expansion around the first zero of the \textit{Jacobi function} cd reads
\begin{align}
  \textrm{cd}(u,\ell) = -\big(u - F(\pi/2,\ell)\big) + \mathcal O\Big(\big(u - F(\pi/2,\ell)\big)^3\Big)\>.
\end{align}
Hence to linear order we find
\begin{align}
   \sqrt{1 - \frac{k^2}{2}} &= \sqrt{1 + \frac{\rho_\text{eff}}{M^2\mu_\text{eff}^2}} \nonumber\\
   &\times\bigg|\mu_\text{eff}\,\sqrt{1 - k^2/2}\,d - F\bigg(\frac{\pi}{2},\frac{|k|/\sqrt2}{\sqrt{1 - k^2/2}}\bigg)\bigg|\>,
\end{align}
respectively
\begin{align}
   \mu_\text{eff}\,d &= \frac{1}{\displaystyle\sqrt{1 - \frac{k^2}{2}}}\,F\bigg(\frac{\pi}{2},\frac{|k|/\sqrt2}{\sqrt{1 - k^2/2}}\bigg) \pm \frac{1}{\displaystyle\sqrt{1 + \frac{\rho_\text{eff}}{M^2\mu_\text{eff}^2}}} \nonumber\\
   &\simeq \frac{1}{\displaystyle\sqrt{1 - \frac{k^2}{2}}}\,F\bigg(\frac{\pi}{2},\frac{|k|/\sqrt2}{\sqrt{1 - k^2/2}}\bigg)\>.
\end{align}
Since
\begin{align}
   \frac{1}{\displaystyle\sqrt{1 - \frac{k^2}{2}}}\,F\bigg(\frac{\pi}{2},\frac{|k|/\sqrt2}{\sqrt{1 - k^2/2}}\bigg) \in [\pi/2,\infty)\>,
\end{align}
we have
\begin{align}
   \mu_\text{eff}\,d \gtrsim \frac{\pi}{2}\>.
\end{align}
In the limit $\rho_\text{eff}\to\infty$ the inequality above becomes rigorous, viz. for $\mu_\text{eff}\,d<\pi/2$ no solution can exist in this limit. This result has been anticipated in \cite{Upadhye:2012rc}, where also the particular solution without any nodes has been approximated by an iteration procedure. Furthermore, in \cite{Brax:2014zta}, the zero-node solution has been obtained  in the approximation of vanishing field values inside the mirrors.

In this work, exact solutions are presented for the first time both inside and outside the mirrors. This is important for weak couplings, viz. for large values of $M$ as well as large values of $\mu$. In those cases the field inside the mirrors is less suppressed. Another novelty, not anticipated before, is the existence of a whole set of solutions for larger values of $d$ due to the periodicity of the \textit{Jacobi function}.
These solutions will be instrumental in analyzing the results of neutron interferometry experiments where the integral of the field across the space between two mirrors measures the phase shift that a neutron experiences on its path to the detectors and also for Casimir experiments and the corresponding calculation of the "Casimir force" induced by the symmetron field. The precise calculations of these effects are left for future work.

\section{2 Mirrors: Anti-Symmetric Solution}\label{sec:4}

In this section, we treat again the case of two parallel infinitely thick mirrors separated at distance $2d$ in $z$-direction, with $z=0$ being the center between the two mirrors. But here, we consider anti-symmetric solutions only.

\subsection{"Broken Symmetry Phase"}

Here, we consider the case of low density $\rho_b < M^2\mu^2$ as between the mirrors and take $z_0=0$. Due to anti-symmetry the field has to vanish there ($\phi_0 = 0$) yielding
\begin{align}
   \frac{1}{2}\left(\frac{d\phi}{dz}\right)^2 - \frac{1}{2}\,\phi_0'^2 = V_\text{eff}(\phi)\>.
\end{align}
where $\displaystyle\phi_0' := \frac{d\phi}{dz}\Big|_{z=0}$. Without loss of generality we take $\phi_0' > 0$, then for $-d\leq z \leq d$ we have
\begin{align}
   \frac{d\phi}{dz} = \pm\sqrt{\phi_0'^2 + 2\,V_\text{eff}(\phi)}\>,
\end{align}
and since $V_\text{eff}(\phi) \leq 0$ this gives a real solution only for $\phi_0'^2 \geq -2\,V_\text{eff}(\phi)$. We have
\begin{align}
   \int_0^{\phi(z)}\frac{d\phi}{\sqrt{\phi_0'^2 - \mu_\text{eff}^2\,\phi^2 + \lambda/2\,\phi^4}} = \pm z\>.
\end{align}
Defining $\tilde\mu^{(\pm)}:=\sqrt{\mu_\text{eff}^2 \pm \sqrt{\mu_\text{eff}^4 - 2\lambda\phi_0'^2}}$ we find
\begin{align}
   \pm |\phi_0'|z &= \frac{\sqrt2\,|\phi_0'|}{\tilde\mu^{(+)}}\int_0^{\tilde\Phi(z)}\frac{d\tilde\phi}{\displaystyle\sqrt{\big(1 - \tilde\phi^2\big)\bigg(1 - \frac{\tilde\mu^{(-)2}}{\tilde\mu^{(+)2}}\,\tilde\phi^2\bigg)}}  \nonumber\\
   &= \frac{\sqrt2\,|\phi_0'|}{\tilde\mu^{(+)}}\,F\bigg(\arcsin\big(\tilde\Phi(z)\big),\frac{\tilde\mu^{(-)}}{\tilde\mu^{(+)}}\bigg)\>,
\end{align}
and obtain
\begin{align}
   \phi(z) = \frac{\sqrt2\,|\phi_0'|}{\tilde\mu^{(+)}}\,\sin\bigg\{F^{-1}\bigg[\pm \frac{\tilde\mu^{(+)}}{\sqrt2}\,z,\frac{\tilde\mu^{(-)}}{\tilde\mu^{(+)}}\bigg]\bigg\}\>.
\end{align}
With the \textit{Jacobi elliptic function}
\begin{align}
   \textrm{sn}(u,\ell) &= \sin\big(F^{-1}(u,\ell)\big)\>,
\end{align}
and
\begin{align}
   \textrm{sn}(-u,\ell) &= -\textrm{sn}(u,\ell)\>,
\end{align}
we can write this as
\begin{align}
   \phi(z) = \pm\frac{\sqrt2\,|\phi_0'|}{\tilde\mu^{(+)}}\,\textrm{sn}\bigg\{\frac{\tilde\mu^{(+)}}{\sqrt2}\,z,\frac{\tilde\mu^{(-)}}{\tilde\mu^{(+)}}\bigg\}\>.
\end{align}

\subsection{"Symmetric Phase"}

Again, we can read off the solution inside the mirrors directly from the corresponding solution in the 1-mirror case
\begin{align}
   &\phi(z) = \sqrt{\frac{2\rho_\text{eff}}{\lambda}}\frac{1}{M}\frac{z}{|z|}\nonumber\\
   &\times\frac{1}{\displaystyle\sinh\bigg(\frac{\sqrt{\rho_\text{eff}}}{M}\,(|z| - d) + \sinh^{-1}\Big(\sqrt{\frac{2\rho_\text{eff}}{\lambda}}\frac{1}{M\phi_d}\Big)\bigg)}\>.
\end{align}

\subsection{Boundary Conditions}

Using the boundary conditions at the surface of the mirror
\begin{align}
   \frac{1}{2}\left(\frac{d\phi}{dz}\right)^2\bigg|_{z=d_-} = \frac{1}{2}\left(\frac{d\phi}{dz}\right)^2\bigg|_{z=d_+}\>,
\end{align}
we find
\begin{align}
  \frac{1}{2}\,\phi_0'^2 - \frac{\mu_\text{eff}^2}{2}\,\phi_d^2 + \frac{\lambda}{4}\,\phi_d^4 = \frac{\rho_\text{eff}}{2M^2}\,\phi_d^2 + \frac{\lambda}{4}\,\phi_d^4\>,
\end{align}
or
\begin{align}
  |\phi_d| = \frac{|\phi_0'|}{\displaystyle\mu_\text{eff}\,\sqrt{1 + \frac{\rho_\text{eff}}{M^2\mu_\text{eff}^2}}}\>.
\end{align}
Using this in the second boundary condition
\begin{align}
   \phi(d_-) = \phi(d_+)\>,
\end{align}
gives
\begin{align}
   \tilde\mu^{(+)} = \sqrt2\,\mu_\text{eff}\,\sqrt{1 + \frac{\rho_\text{eff}}{M^2\mu_\text{eff}^2}}\,\bigg|\textrm{sn}\bigg\{\frac{\tilde\mu^{(+)}}{\sqrt2}\,d,\frac{\tilde\mu^{(-)}}{\tilde\mu^{(+)}}\bigg\}\bigg|\>.
\end{align}
The solution of the equation above gives the absolute value of $\phi_0'$. It will give all solutions for which $\phi'(d_-) = \pm\phi'(d_+)$, hence one still has to extract the subset of solutions satisfying $\phi'(d_-) = \phi'(d_+)$.

\subsection{Final Solution}

Finally, we obtain the solution
\begin{align}
 \phi(z) &= \pm\Theta(d - |z|)\,\frac{\sqrt2\,|\phi_0'|}{\tilde\mu^{(+)}}\,\textrm{sn}\bigg\{\frac{\tilde\mu^{(+)}}{\sqrt2}\,z,\frac{\tilde\mu^{(-)}}{\tilde\mu^{(+)}}\bigg\} \nonumber\\
   &\quad\pm \Theta(|z| - d)\,\sqrt{\frac{2\rho_\text{eff}}{\lambda}}\frac{1}{M}\frac{z}{|z|}\nonumber\\
   &\times\frac{\mathfrak{sgn_A}}{\displaystyle\sinh\bigg(\frac{\sqrt{\rho_\text{eff}}}{M}\,(|z| - d) + \sinh^{-1}\Big(\sqrt{\frac{2\rho_\text{eff}}{\lambda}}\frac{1}{M\phi_d}\Big)\bigg)}\>.
\end{align}
where
\begin{align}
  \mathfrak{sgn_A} := \frac{\displaystyle\textrm{sn}\bigg\{\frac{\tilde\mu^{(+)}}{\sqrt2}\,d,\frac{\tilde\mu^{(-)}}{\tilde\mu^{(+)}}\bigg\}}{\displaystyle\bigg|\textrm{sn}\bigg\{\frac{\tilde\mu^{(+)}}{\sqrt2}\,d,\frac{\tilde\mu^{(-)}}{\tilde\mu^{(+)}}\bigg\}\bigg|}\>,
\end{align}
and $|\phi_0'|$ is solution of
\begin{align}\label{SSA}
   \tilde\mu^{(+)} = \sqrt2\,\mu_\text{eff}\,\sqrt{1 + \frac{\rho_\text{eff}}{M^2\mu_\text{eff}^2}}\,\bigg|\textrm{sn}\bigg\{\frac{\tilde\mu^{(+)}}{\sqrt2}\,d,\frac{\tilde\mu^{(-)}}{\tilde\mu^{(+)}}\bigg\}\bigg|\>,
\end{align}
and
\begin{align}
  \phi_d = \pm\frac{|\phi_0'|}{\displaystyle\mu_\text{eff}\,\sqrt{1 + \frac{\rho_\text{eff}}{M^2\mu_\text{eff}^2}}}\>.
\end{align}
As mentioned before, Eq.~(\ref{SSA}) will give all solutions for which $\phi'(d_-) = \pm\phi'(d_+)$, hence one still has to extract the subset of solutions satisfying $\phi'(d_-) = \phi'(d_+)$.

\begin{figure}
\begin{center}
\epsfxsize=8.5 cm \epsfysize=5.2 cm {\epsfbox{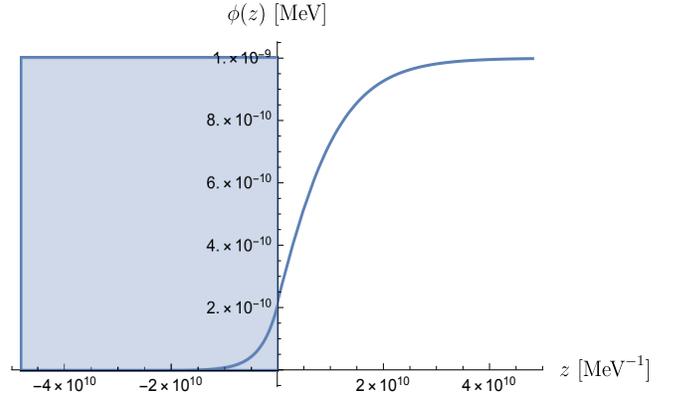}}
\end{center}
\caption{The 1 mirror solution is depicted for the parameters of Tab.~\ref{table:APE}. One can see that the condition $\phi/M\ll1$, necessitated by Eq.~(\ref{cf}), is indeed satisfied in this case.}
\label{Fig4}
\end{figure}

For most situations of interest, where $\rho_\text{eff} \gg M^2\mu_\text{eff}^2$, the left-hand side of Eq.~(\ref{SSA}) is typically small, while the factor $\sqrt{1 + \frac{\rho_\text{eff}}{M^2\mu_\text{eff}^2}}$ is comparably large.
Hence, a solution demands that the \textit{Jacobi function} sn should be close to its zero, which is the case for a vanishing first argument. For the first zero this leads to solutions obeying
$\phi'(d_-) = -\phi'(d_+)$ rather than $\phi'(d_-) = \phi'(d_+)$. The Taylor expansion around the second zero of the \textit{Jacobi function} sn reads
\begin{align}
  \textrm{sn}(u,\ell) = -\big(u - 2 F(\pi/2,\ell)\big) + \mathcal O\Big(\big(u - 2 F(\pi/2,\ell)\big)^3\Big)\>.
\end{align}
Hence to linear order we find
\begin{align}
   \tilde\mu^{(+)} = \sqrt2\,\mu_\text{eff}\,\sqrt{1 + \frac{\rho_\text{eff}}{M^2\mu_\text{eff}^2}}\,\bigg|\frac{\tilde\mu^{(+)}d}{\sqrt2} - 2 F\bigg(\frac{\pi}{2},\frac{\tilde\mu^{(-)}}{\tilde\mu^{(+)}}\bigg)\bigg|\>,
\end{align}
respectively
\begin{align}
   \tilde\mu^{(+)}d &= \sqrt2\,2 F\bigg(\frac{\pi}{2},\frac{\tilde\mu^{(-)}}{\tilde\mu^{(+)}}\bigg) \pm \frac{\tilde\mu^{(+)}}{\displaystyle\mu_\text{eff}\,\sqrt{1 + \frac{\rho_\text{eff}}{M^2\mu_\text{eff}^2}}} \nonumber\\
   &\simeq \sqrt2\,2 F\bigg(\frac{\pi}{2},\frac{\tilde\mu^{(-)}}{\tilde\mu^{(+)}}\bigg)\>.
\end{align}
Since
\begin{align}
   \sqrt2\,2 F\bigg(\frac{\pi}{2},\frac{\tilde\mu^{(-)}}{\tilde\mu^{(+)}}\bigg) \in [\sqrt2\,\pi,\infty),
\end{align}
and
\begin{align}
   \tilde\mu^{(+)} \in [\mu_\text{eff},\sqrt2\,\mu_\text{eff}]\>,
\end{align}
we have
\begin{align}
   \mu_\text{eff}\,d \gtrsim \pi\>.
\end{align}
In the limit $\rho_\text{eff}\to\infty$ the inequality above becomes rigorous, viz. for $\mu_\text{eff}\,d<\pi$ no solution can exist in this limit. Here, $d$ must be twice as large as in the case of the symmetric solution, which is intuitively clear.

\section{Discussion of the 2 Mirror Solutions}\label{sec:5}

An important question to consider is whether the symmetric and anti-symmetric solutions exhaust the set of possible solutions or whether there are still others without definite (anti-)symmetry? One may argue that a solution without any symmetry must satisfy two boundary conditions (continuity of $\phi(z)$ and $\phi'(z)$) for the left boundary and a set of two different boundary conditions for the right boundary. If we start with a symmetric solution, then all four boundary conditions are satisfied. Moving the right mirror to generate another solution without symmetry, we keep the parameter $k$ fixed (in order to still satisfy the left boundary conditions) and vary the distance $d^R$  to the right mirror. For a given value of $k$, this leads to a resonance equation for $d^R$ whose solution is either the original one $d^R=d^L$ or correspond to new solutions. One may construct new solutions which are either symmetric again or anti-symmetric but will have a different number of nodes from the solution we started with. This depends on the value of $d_R$. A more thorough discussion of the possible existence of other solutions together with  the symmetric and anti-symmetric ones is left for future work.
On the other hand, it is not obvious how the symmetric and anti-symmetric solutions should be expected to appear in an experimental setup. In particular whether the solutions are stable or not and which one is the stablest. One can expect that the solution with the fewest number of zeros between the two mirrors should be the ground state, although we have no proof of that assertion and further study is certainly necessary to tackle this important question.

\section{A Particular Example}\label{sec:6}

Let us focus on a particular example for the parameters given in Tab.~\ref{table:APE} for 2 mirrors. This illustrates the different cases presented in the previous sections and provides an example of multiple solutions.  The corresponding 1 mirror solution with $k = 0.214967$ is depicted in Fig.~\ref{Fig4}.
\begin{table}[!ht]
\centering
\addtolength{\tabcolsep}{2pt}
\renewcommand{\arraystretch}{1.5}
\begin{tabular}{|c||c|c|c|}
  \hline
  Mode & $E$ [MeV]  & $k$ & $|\phi_0'|$ [MeV$^2$] \\
  \hline\hline
$0^+$ & $-1.76369\times10^{-28}$  & $0.997142$ & \\
  \hline
$1^-$ & $-8.30488\times10^{-29}$  && $6.9018\times10^{-20}$ \\
  \hline
$2^+$ & $-9.36839\times10^{-30}$  & $0.451367$ &\\
  \hline
\end{tabular}
\caption{Values of energy $E$ are given for the three solutions for values $\rho_\text{eff} = 1.082\times10^{-5}$ MeV$^4$, $M=10^7$ MeV, $\mu_\text{eff}=10^{-10}$ MeV, $\lambda = 10^{-2}$ and mirror distance $d=9.5\times10^{-3}$ m.}
\label{table:APE}
\end{table}

\begin{figure}
\begin{center}
\epsfxsize=8.5 cm \epsfysize=5.2 cm {\epsfbox{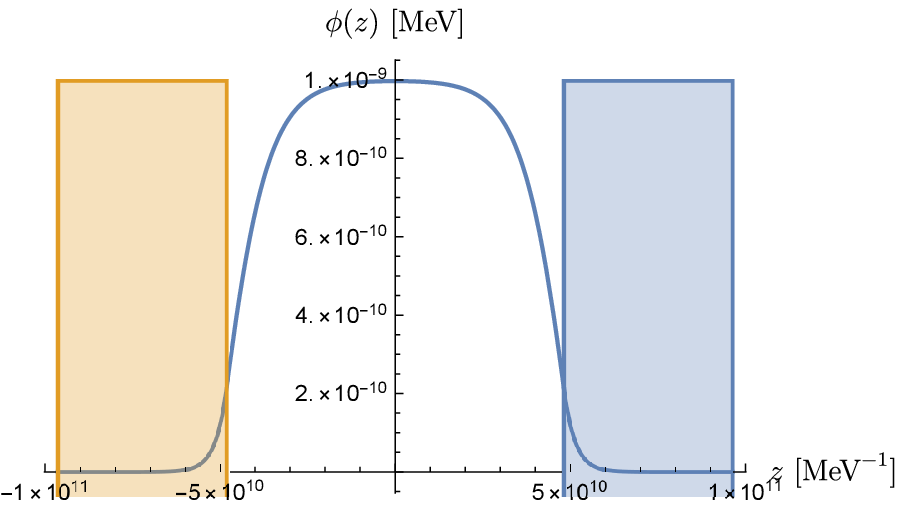}}
\epsfxsize=8.5 cm \epsfysize=5.2 cm {\epsfbox{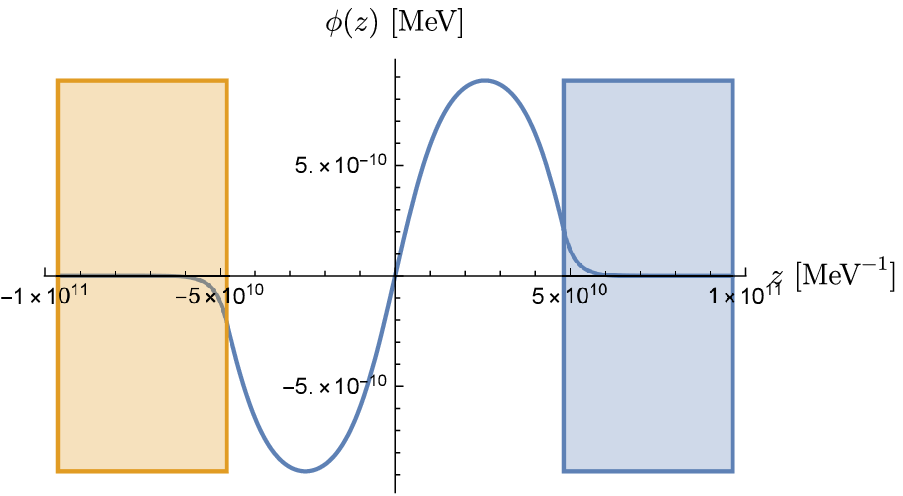}}\\
\epsfxsize=8.5 cm \epsfysize=5.2 cm {\epsfbox{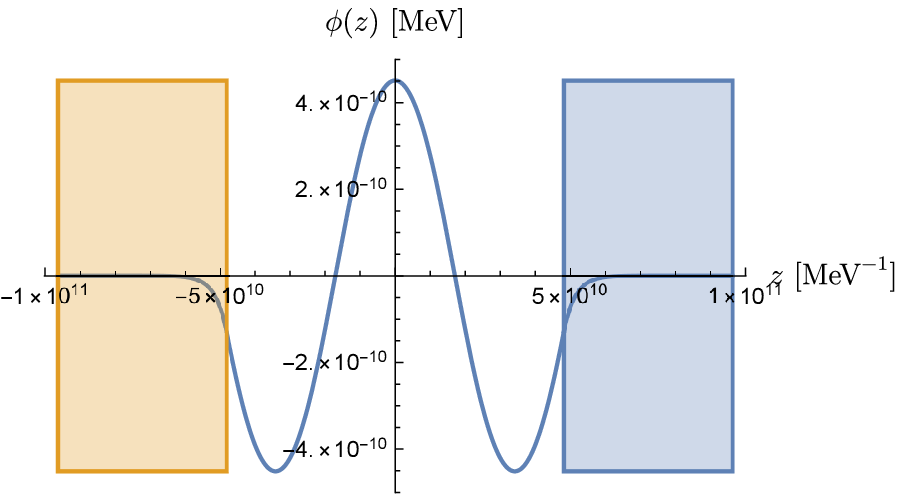}}
\end{center}
\caption{The field profiles for the three solutions from Tab.~\ref{table:APE} are depicted. Again, one can see that the condition $\phi/M\ll1$, necessitated by Eq.~(\ref{cf}), is indeed satisfied in this case.}
\label{Fig5}
\end{figure}

The energies
\begin{align}
  E = \int_{-\infty}^\infty dz\,\mathcal H(z)\>,
\end{align}
where the Hamiltonian $\mathcal H(z)$ is given by Eq.~(\ref{Ham}), for the complete set of three solutions are also given in Tab.~\ref{table:APE} and the corresponding field profiles are depicted in Fig.~\ref{Fig5}. We denote solutions, viz. modes by their number of nodes $0,1,2$ and an upper index $+$ to denote symmetry as well as $-$ to denote anti-symmetry. From Tab.~\ref{table:APE} one can see that the $0^+$ mode has lowest energy, $1^-$ a higher one and $2^+$ the highest energy. The energy level $E=0$ would correspond to the vanishing solution $\phi(z)=0$. These three solutions exhaust the spectrum of possible solutions for these parameter values and  exhibits clearly how anti-symmetric and multiple node solutions are a crucial part of the complete set of solutions.

Intuitively, we expect the energies to increase with the order of the nodes not only in this particular example but also in general. Clearly, depending on the parameter values and distance between the mirrors one can have an arbitrary number of nodes.

\section{Symmetron induced Frequency Shift in \textit{q}BOUNCE}\label{sec:7}

In this section we derive limits obtained from the \textit{q}BOUNCE experiment \cite{Abele:2009dw, Jenke:2011zz, Jenke:2014yel} using the exact solutions obtained herein. In this experiment, ultra-cold neutrons are dropped in earth's gravitational potential and reflected by a neutron mirror,  which has been reported for the first time in \cite{Nesvizhevsky:2002ef}. The energy eigenstates are discrete and allow to apply the method of resonance spectroscopy. The basic setup is described in \cite{Jenke:2011zz}. In the most recent version of the experiment, Rabi-spectroscopy has been realized with energy resolution 3$\times$10$^{-15}$ peV \cite{Cronenberg:2017aa}.

The experimental setup is such, that ultra-cold neutrons pass three regions, while being reflected on polished glass mirrors. In \cite{Cronenberg:2017aa}, the resonance spectroscopy transition between the energy ground state $E_1 = 1.40672$ peV and the excited states $E_3 = 3.32144$ peV as well as $E_4 = 4.08321$ peV have been demonstrated. First, the neutrons pass region I which acts as a state selector for the ground state $|1\rangle$ having energy $E_1$. A polished mirror at the bottom and a rough absorbing scatterer on top at a height of about 20 $\mu$m serve to select the ground state. Neutrons in higher, unwanted states are scattered out of the system. This region has a length of 15 cm. Subsequently, in region II, a horizontal mirror performs harmonic oscillations with a tunable frequency $\omega$, which drives the system into a coherent superposition of ground and excited states. The length of this region is 20 cm. Finally, region III is identical to the first region and hence  acts again as a ground state selector.
The quantum-mechanical description of a neutron above a mirror in the gravitational potential is given by the Schr\"odinger equation \cite{Westphal:2006dj}.
After separation into free transversal and bound vertical states
\begin{align}
   \Psi_n^{(0)}(\textbf{x},t) = \frac{e^{\frac{i}{\hbar}(p_\perp\cdot x_\perp - E_\perp t)}}{2\pi\hbar v_\perp}\,\psi_n^{(0)}(z)\,e^{-\frac{i}{\hbar}E_n t}\>,
\end{align}
it reads
\begin{align}\label{SEQ}
   -\frac{\hbar^2}{2m}\frac{\partial^2\psi_n(z)}{\partial z^2} + mgz\,\psi_n(z) = E_n\psi_n(z)\>.
\end{align}
The characteristic length scale
\begin{align}
  z_0 = \sqrt[3]{\frac{\hbar^2}{2m^2g}} = 5.87\,\mu\text{m}\>,
\end{align}
and energy scale $E_0$ = $((\hbar^2 mg^2)/2)^{1/3}$ are given by the mass $m$ of the neutron and the acceleration of the earth $g$. With the substitution
\begin{align}
  \sigma = \sqrt[3]{\frac{2m^2g}{\hbar^2}}\left(z - \frac{E_n}{mg}\right) \equiv \frac{z - z_n}{z_0}\>,
\end{align}
Eq.~(\ref{SEQ}) becomes
\begin{align}
   \frac{d^2\tilde\psi_n(\sigma)}{d\sigma^2} - \sigma\,\tilde\psi_n(\sigma) = 0\>,
\end{align}
which is Airy's equation.

From the effective symmetron potential
\begin{align}
   V_\text{eff}(\phi) = \frac{1}{2}\left(\frac{\rho}{M^2} - \mu^2\right)\phi^2 + \frac{\lambda}{4}\,\phi^4\>,
\end{align}
we can deduce the semi-classical neutron-symmetron coupling as
\begin{align}\label{nsc}
   V_\text{eff} = \frac{1}{2}\frac{m}{M^2}\,\psi^*\psi\,\phi^2\>.
\end{align}
There are some subtleties involved here, which will be discussed in the Appendix.
The corresponding quantum mechanical perturbation potential is given by
\begin{align}\label{symmQMP}
   \textrm{V} = \frac{1}{2}\frac{m}{M^2}\,\phi^2\>,
\end{align}
and leads to a resonance frequency shift  (see e.g. \cite{landau1991quantenmechanik})
\begin{align}
   \delta E_{mn}^{(1)} &\equiv E_m^{(1)} - E_n^{(1)} \nonumber\\
   &= \frac{1}{2}\frac{m}{M^2}\int_{-\infty}^\infty dz\,\Big(\big|\psi_m^{(0)}(z)\big|^2 - \big|\psi_n^{(0)}(z)\big|^2\Big)\,\phi(z)^2\>.
\end{align}
Likewise, the first order correction to the wavefunctions reads (see e.g. \cite{landau1991quantenmechanik})
\begin{align}
   &\psi_n^{(1)}(z,t) = \frac{1}{2}\frac{m}{M^2} \nonumber\\
   &\times\sum_{m=1\setminus n}^\infty\frac{\displaystyle\int_{-\infty}^\infty dz'\,\psi_m^{(0)*}(z')\,\psi_n^{(0)}(z')\,\phi(z')^2}{E_n^{(0)} - E_m^{(0)}}\,\psi_m^{(0)}(z)\,e^{-\frac{i}{\hbar}E_n t}\>.
\end{align}
Hence, the correction to the density $\displaystyle\varrho_n^{(0)}(z)=\psi_n^{(0)*}(z)\psi_n^{(0)}(z)$ to first order is given by
\begin{align}
   &\varrho_n^{(1)}(z) = 2\,\mathfrak{Re}\big(\psi_n^{(0)*}(z)\psi_n^{(1)}(z)\big)  \nonumber\\
   &= \frac{m}{M^2}\,\mathfrak{Re}\sum_{m=1\setminus n}^\infty\frac{\displaystyle\int_{-\infty}^\infty dz'\,\psi_m^{(0)*}(z')\,\psi_n^{(0)}(z')\,\phi(z')^2}{E_n^{(0)} - E_m^{(0)}}  \nonumber\\
   &\quad\times\psi_n^{(0)*}(z)\psi_m^{(0)}(z)\>.
\end{align}
where $\mathfrak{Re}$ denotes the real part.

In the 1-mirror case
the unperturbed normalized wavefunction for $z>0$ reads (see e.g. \cite{pitschmann2017qbounce})
\begin{align}
  \psi_n^{(0)}(z) = C_n^{(1)}\text{Ai}\Big(\frac{z - z_n}{z_0}\Big)\>,
\end{align}
with normalisation
\begin{align}
  C_n^{(1)} = \frac{1}{\sqrt{z_0}\,\text{Ai}'\Big(-\displaystyle\frac{z_n}{z_0}\Big)}\>,
\end{align}
and $\displaystyle z_n = \frac{E_n}{mg}$.
Outside this region the wavefunction vanishes. The first few energy levels are given in table~\ref{table:EL}.
\begin{table}[!ht]
\centering
\addtolength{\tabcolsep}{2pt}
\renewcommand{\arraystretch}{1.5}
\begin{tabular}{|c|c|}
	\hline
	State & Energy [peV] \\
	\hline\hline
	 $|1\rangle$ & $E_1=1.40672$ \\
	\hline	
	 $|2\rangle$ &  $E_2=2.45951$ \\
	\hline	
	 $|3\rangle$ &  $E_3=3.32144$ \\
	\hline	
	 $|4\rangle$ &  $E_4=4.08321$ \\
	\hline	
	 $|5\rangle$ &  $E_5=4.77958$ \\
	\hline
	$|6\rangle$ &  $E_6=5.42846$ \\
	\hline
\end{tabular}
\caption{Values of the  energy of the lowest six states for the neutrons in the terrestrial gravitational field.}
\label{table:EL}
\end{table}
For a single mirror extending to $z \leq 0$ we can use Eq.~(\ref{FS1M}) and obtain for the resonance frequency shift
\begin{align}\label{RFS1}
   \delta E_{mn}^{(1)} &= \frac{1}{2}\frac{m_N}{M^2}\frac{\mu_\text{eff}^2}{z_0\lambda}\int_0^\infty dz\,\tanh\!\Big(\frac{\mu_\text{eff} z}{\sqrt2} + \tanh^{-1}k\Big)^2 \nonumber\\
   &\quad\times\Bigg\{\frac{\displaystyle\text{Ai}\Big(\frac{z - z_m}{z_0}\Big)^2}{\text{Ai}'\Big(-\displaystyle\frac{z_m}{z_0}\Big)^2} - \frac{\displaystyle\text{Ai}\Big(\frac{z - z_n}{z_0}\Big)^2}{\text{Ai}'\Big(-\displaystyle\frac{z_n}{z_0}\Big)^2}\Bigg\}\>.
\end{align}

It is straightforward to find all the corresponding expressions in the 2-mirror case. These expressions are very elaborate in their full detail and hence we will refrain from reproducing them herein.

In Table \ref{table1} we summarize the resonance frequency shifts for a large range of symmetron parameters for the case of a single mirror.
\begin{table}[ht]
\centering
\renewcommand{\arraystretch}{1.5}
\begin{tabular}{|l||c|l|}
  \hline
  $\delta E_{14}^{(1)}$ [eV] 	&	$M$ [MeV]		& 	\quad $\lambda$\\
 \hline\hline
$4.49795\times 10^{-14}$ 		&  	$1$ 		& 	$10^{-10}$  \\
\hline
$4.49795\times 10^{-20}$ 		& 	$1$ 		& 	$10^{-4}$  \\
\hline
$4.49795\times 10^{-26}$ 		&  	$1$ 		& 	$10^2$  \\
\hline
$4.51053\times 10^{-20}$ 		&  	$10^3$ 	& 	$10^{-10}$  \\
\hline
$4.51053\times 10^{-26}$ 		& 	$10^3$ 	& 	$10^{-4}$  \\
\hline
$4.51053\times 10^{-32}$ 		&  	$10^3$ 	& 	$10^2$  \\
\hline
$5.75763\times 10^{-24}$ 		&  	$10^5$ 	& 	$10^{-10}$  \\
\hline
$5.75763\times 10^{-30}$ 		&  	$10^5$ 	& 	$10^{-4}$  \\
\hline
$5.75763\times 10^{-36}$ 		& 	$10^5$ 	& 	$10^2$  \\
\hline
\end{tabular}
\caption{$\delta E_{41}^{(1)}$ for $1$ mirror, viz. Eq.~(\ref{RFS1}), for the values $\rho = 2.51$ g/cm$^3$ and $\mu_\text{eff} = 10^{-3}$ meV. Notice larger values of $\lambda$ and $M$ lead to smaller energy shifts. The current experimental bound at the $10^{-15}$ eV level leads to constraints on parameters which can be easily extracted using our analysis \cite{Cronenberg:2017aa}. For instance for $M=1$ MeV, we find that typically one can expect that  $\lambda \gtrsim 10^{-10}$. The exact excluded regions in parameter space, where the screening of the neutron is taken into account, can be found in \cite{Cronenberg:2017aa} (see the Appendix for a discussion).}
\label{table1}
\end{table}
Notice that larger values of $\lambda$ give smaller energy shifts for a given  $\mu_\text{eff}$. Similarly, increasing $M$ leads to smaller deviations. The whole parameter space of symmetrons can therefore be efficiently constrained using the exact results obtained herein.
A sophisticated $\chi^2$ analysis has been carried out in parallel to this work \cite{Cronenberg:2017aa}, where the solutions obtained herein for a 1 mirror setup are used to extend the exclusion region of the symmetron parameter space.

\section{Conclusion}\label{sec:8}

We have derived exact analytical solutions to the symmetron field theory in the presence of a one or two mirror system. The one dimensional equations of motion have been integrated in each case.
Surprisingly, in the two mirror case instead of a unique solution for a given environment and boundary conditions, we have obtained a discrete set of solutions with increasing number of nodes and energies. We have derived bounds for the \textit{q}BOUNCE experiment and found agreement with those obtained in the study carried out in parallel to this work \cite{Cronenberg:2017aa}. We have summarized the resonance frequency shift for a large range of symmetron parameters in the case of a single mirror. The solutions that we have obtained herein, and in particular their stability, should be the subject of further studies as they will play a crucial role in the analysis of neutron interferometry experiments searching for symmetrons and for the calculation of the symmetron field induced "Casimir force" in the CANNEX experiment.

\acknowledgments

We thank Hartmut Abele, Guillaume Pignol and Ren\'e Sedmik for fruitful discussions. This work is supported in part by the EU Horizon 2020 research and innovation programme under the Marie-Sklodowska grant No. 690575. This article is based upon work related to the COST Action CA15117 (CANTATA) supported by COST (European Cooperation in Science and Technology).

\appendix

\section{Symmetron Field of a Neutron}\label{sec:A}

Since neutrons are used in the search for symmetrons in \textit{q}BOUNCE experiments and neutron interferometry, it is important to understand their interaction with the symmetron. For a certain parameter regime this interaction between neutron and symmetron becomes strong. In this case, the neutron affects the background symmetron field as generated by the mirrors of the experimental setup in a non-negligible way, which in turn weakens the effect on the neutron, viz. screening of the neutron sets in.
Since we treat the symmetron as a classical field theory a consistent description of its coupling to a quantum mechanical system is beyond our reach. Therefore, we employ a semi-classical treatment in which the neutron's probability distribution times its mass acts as the source of the symmetron as defined in Eq.~(\ref{nsc}).
In the following we will use a pragmatic approach, which follows what has been done in the literature so far \cite{Burrage:2014oza, Burrage:2015lya, Burrage:2014daa, Burrage:2016rkv, Burrage:2016lpu, Burrage:2017shh, Burrage:2016xzz}.

A first approach to defining the mass density is analogous to "semi-classical gravity", which has been introduced by M{\o}ller \cite{moller1962theories} and Rosenfeld \cite{ROSENFELD1963353}, and where the operator-valued stress-energy tensor is replaced by an expectation value. Its non-relativistic Newtonian approximation, the so-called "Schr\"odinger-Newton equation", was introduced by Di\'osi \cite{Diosi:2014ura} and Penrose \cite{Penrose:1998dg}. This equation employs the probability distribution times the mass as the source of the gravitational potential, i.e. considering that the square modulus of the wave function can be used as tracing the mass density of a particle.  The "Schr\"odinger-Newton equation" has been applied to single particles by Moroz, Penrose and Tod \cite{Moroz:1998dh}, although it can only be understood in the Hartree approximation for a large ensemble of particles, see \cite{Bahrami:2014gwa} for  a summary of the issues related to this equation.
In another approach, which is commonly used in the literature, one may consider that the neutron has a well-defined size provided by the quark-gluon dynamics and that in our non-relativistic treatment of the neutron at energies well-below the QCD scale this size of the order of $1$ fm provides a reasonable description for its screening. Clearly, a rigorous treatment of this issue and distinguishing both approaches is beyond the scope of this paper. In \cite{Cronenberg:2017aa}, we illustrate the two cases and stress that more work needs to be done to derive completely rigorous bonds for the coupling of the symmetron to neutrons.

For a quantitative account we treat the neutron as a classical sphere. As a conservative case we take as its diameter $2R_n = 1$ fm, corresponding to $5\times 10^{-3}$ MeV$^{-1}$ in natural units, and nuclear density $\rho_n = 1.53\times10^{10}$ MeV$^{4}$. This is the most commonly used manner of treating the screening of the neutron. On the other hand, if the mass distribution were related to the size of its wave function, the neutron would be approximately described as being distributed over a sphere of diameter of order $z_0 = 5.87\,\mu\text{m}$, which is the typical length scale for its vertical wavefunction in the \textit{q}BOUNCE experiment and leads to a  much weaker screening.

As a result, we have to solve the field equation for a static massive sphere with radius $R$
\begin{align}
   \frac{d^2\phi}{dr^2} + \frac{2}{r}\frac{d\phi}{dr} = \left(\frac{\rho}{M^2} - \mu^2\right)\phi + \lambda\,\phi^3\>.
\end{align}
The boundary conditions are
\begin{align}
   \phi'(0) &= 0\>, \nonumber\\
   \lim_{r\to\infty}\phi(r) &\to \pm\phi_V\>.
\end{align}
Without loss of generality we take the asymptotic value $+\phi_V$.
\begin{figure}
\begin{center}
\epsfxsize=8.5 cm \epsfysize=5.2 cm {\epsfbox{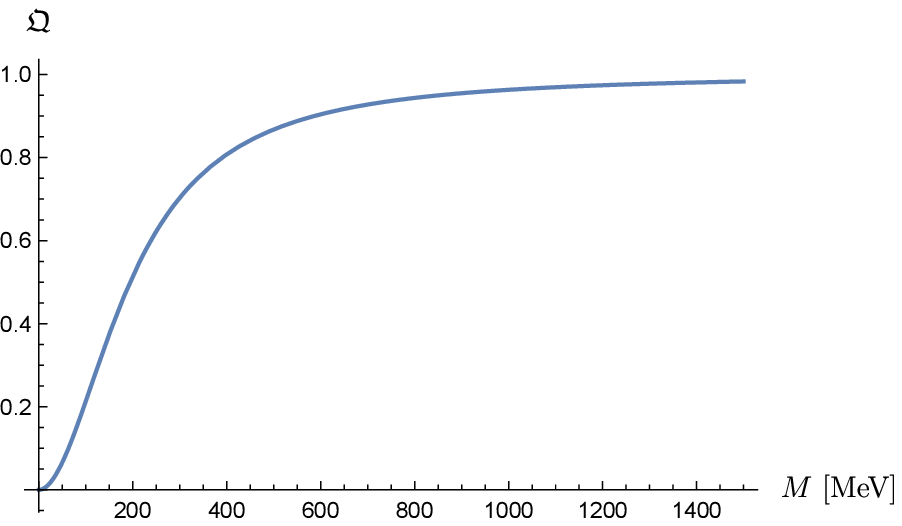}}
\epsfxsize=8.5 cm \epsfysize=5.2 cm {\epsfbox{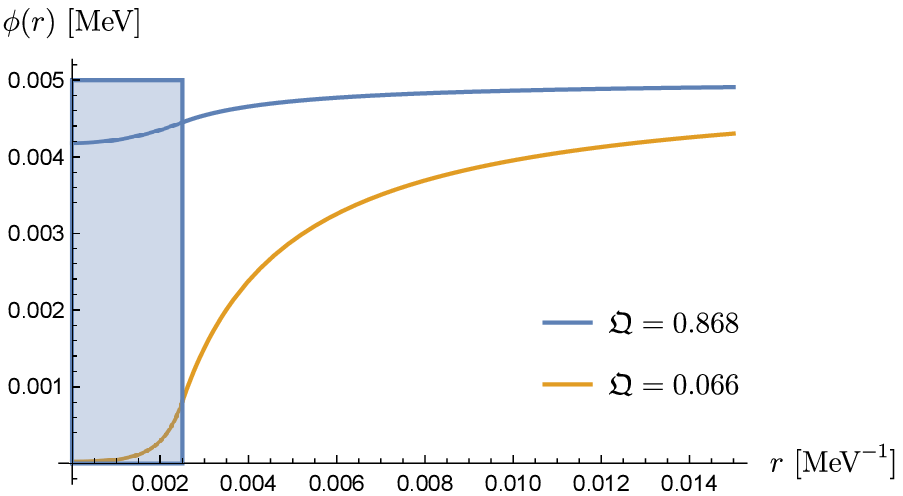}}
\end{center}
\caption{\textit{Top:} Here the \textit{screening charge} $\mathfrak Q$ is plotted as a function of the coupling parameter $M$. The parameters taken are for a neutron, \textit{viz.} $R = 2.5\times10^{-3}$ MeV$^{-1}$, $\rho_n = 1.53\times10^{10}$ MeV$^{4}$ and $\mu = 5\times10^{-8}$ MeV.
\textit{Bottom:} The field profile of a sphere (neutron) is plotted as a function of the radial distance of the center of the sphere. The blue line corresponds to $M = 500$ MeV and the yellow line to $M = 50$ MeV with the other parameters having the same values as those used for the upper plot. The blue square is bounded by the radius of the sphere (neutron) and the vacuum field value $\phi = \phi_V$.}
\label{FigScr}
\end{figure}
Inside the sphere we take $\rho\gg M^2\mu^2$ and since
\begin{align}
   \frac{\displaystyle\left(\frac{\rho}{M^2} - \mu^2\right)\phi}{\displaystyle\lambda\,\phi^3} \simeq \frac{\displaystyle\rho}{\displaystyle M^2\lambda\,\phi^2} \gtrsim \frac{\displaystyle\rho}{\displaystyle M^2\lambda\,\phi_V^2} = \frac{\displaystyle\rho}{\displaystyle M^2\mu^2} \gg 1\>,
\end{align}
we find
\begin{align}\label{eomsi}
   \frac{d^2\phi}{dr^2} + \frac{2}{r}\frac{d\phi}{dr} \simeq \frac{\rho}{M^2}\,\phi\>.
\end{align}
With
\begin{align}
   \phi = \frac{\varphi}{r}\>,
\end{align}
Eq.~(\ref{eomsi}) reads
\begin{align}
   \frac{d^2\varphi}{dr^2} \simeq \frac{\rho}{M^2}\,\varphi\>,
\end{align}
providing the general solution
\begin{align}
   \phi = A\,\frac{e^{\frac{\sqrt\rho}{M}r}}{r} + B\,\frac{e^{-\frac{\sqrt\rho}{M}r}}{r}\>,
\end{align}
with arbitrary constants $A$ and $B$. The solution convergent for $r\to0$ and satisfying the boundary condition $\phi(0) = 0$ is given by $B = -A =: -C/2$
\begin{align}
   \phi(r) = C\,\frac{\displaystyle\sinh\Big({\frac{\sqrt\rho}{M}\,r}\Big)}{r}\>.
\end{align}
Immediately outside the sphere, the potential contribution is very small compared to the kinetic parts and can be approximated by the value close to the vacuum value
\begin{align}
  \frac{\rho}{M^2}\,\phi(R_-) + \lambda\,\phi(R_-)^3 &\gg   - \mu^2\,\phi(R_+) + \lambda\,\phi(R_+)^3 \nonumber\\
  &\simeq 2\mu^2(\phi(R_+) - \phi_V)\>.
\end{align}
Hence, for the outside solution we find
\begin{align}\label{eomso}
   \frac{d^2\phi}{dr^2} + \frac{2}{r}\frac{d\phi}{dr} \simeq 2\mu^2(\phi - \phi_V)\>.
\end{align}
The general solution is given by
\begin{align}
   \phi - \phi_V = E\,\frac{e^{\sqrt2\mu r}}{r} + D\,\frac{e^{-\sqrt2\mu r}}{r}\>,
\end{align}
with arbitrary constants $E$ and $D$. The solution convergent for $r\to\infty$ and satisfying the boundary condition $\lim_{r\to\infty}\phi(r) \to \phi_V$ is given by
\begin{align}
   \phi(r) = \phi_V + D\,\frac{e^{-\sqrt2\mu r}}{r}\>.
\end{align}
The boundary conditions at the surface of the sphere are given by
\begin{align}
   \phi(R_-) &= \phi(R_+)\>, \nonumber\\
   \phi'(R_-) &= \phi'(R_+)\>,
\end{align}
yielding after some algebraical manipulations
\begin{align}
   C &= \phi_V\frac{M}{\sqrt\rho}\frac{1}{\displaystyle\cosh\Big(\frac{\sqrt\rho}{M}R\Big)}\frac{\displaystyle1 + \sqrt2\mu R}{\displaystyle1 + \sqrt2\mu\,\frac{M}{\sqrt\rho}\,\tanh\Big(\frac{\sqrt\rho}{M}R\Big)}\>, \nonumber\\
   D &= - \phi_VR\,e^{\sqrt2\mu R}\,\frac{\displaystyle1 - \frac{M}{\sqrt\rho}\frac{1}{R}\,\tanh\Big(\frac{\sqrt\rho}{M}R\Big)}{\displaystyle1 + \sqrt2\mu\,\frac{M}{\sqrt\rho}\,\tanh\Big(\frac{\sqrt\rho}{M}R\Big)}\>.
\end{align}
Finally, we find the solution
\begin{align}
  \phi(r) = \begin{dcases}
    \phi_V\frac{M}{\sqrt\rho}\frac{1}{\displaystyle\cosh\Big(\frac{\sqrt\rho}{M}R\Big)}\frac{\displaystyle1 + \sqrt2\mu R}{\displaystyle1 + \sqrt2\mu\,\frac{M}{\sqrt\rho}\,\tanh\Big(\frac{\sqrt\rho}{M}R\Big)} \nonumber\\
    \qquad\times\frac{\displaystyle\sinh\Big({\frac{\sqrt\rho}{M}\,r}\Big)}{r}\>,  \quad \text{for } r\leq R\>, \nonumber\\
    \phi_V - \frac{\rho R^3}{3M^2}\frac{\mathfrak Q\,\phi_V}{\displaystyle1 + \sqrt2\mu R}\frac{e^{-\sqrt2\mu (r - R)}}{r}\>, \>\text{for } r\geq R\>,
  \end{dcases}
\end{align}
where we have introduced a \textit{screening charge}, which we define as
\begin{align}
   \mathfrak Q = \big(1 + \sqrt2\mu R\big)\,\frac{3M^2}{\rho R^2}\frac{\displaystyle1 - \frac{M}{\sqrt\rho}\frac{1}{R}\,\tanh\Big(\frac{\sqrt\rho}{M}R\Big)}{\displaystyle1 + \sqrt2\mu\,\frac{M}{\sqrt\rho}\,\tanh\Big(\frac{\sqrt\rho}{M}R\Big)}\>.
\end{align}
With this definition
\begin{align}
  \mathfrak Q \to \begin{dcases}
    0\>, & \quad \text{for \textit{screened} bodies with }R\gg M/\sqrt\rho\>, \nonumber\\
    1\>, & \quad \text{for \textit{unscreened} bodies with }R\ll M/\sqrt\rho\>.
  \end{dcases}
\end{align}

The acceleration of a small test body, which does not disturb the field, in the outer field of the sphere is
 \begin{align}
   \vec a = -\frac{\phi}{M^2}\,\vec\nabla\phi\>.
\end{align}
Asymptotically for large $r$ we find
\begin{align}
   \vec a = -\mathfrak Q\,\phi_V^2\frac{\rho R^3}{3M^4}\frac{\sqrt2\mu}{\displaystyle1 + \sqrt2\mu R}\frac{e^{-\sqrt2\mu r}}{r}\frac{\vec r}{r}\>,
\end{align}
justifying the definition of $\mathfrak Q$ as a \textit{screening charge}, which has to be multiplied to the transition energies. Hence, in order to account for the neutron's screening one has to replace
\begin{align}
  \delta E_{pq} \to \mathfrak Q(\mu,M)\,\delta E_{pq},\label{eq:ScreenedDeltaE}
\end{align}
for the extraction of the experimental limits. This replacement has been carried out in the sophisticated analysis \cite{Cronenberg:2017aa} but for simplicity has been neglected for all bounds derived in this letter.

\newpage

\bibliographystyle{unsrt}
\bibliography{Symmetron}

\end{document}